\documentclass[twocolumn,aps,prl,showpacs,superscriptaddress,preprintnumbers,floatfix,nofootinbib,longbibliography, notitlepage]{revtex4-2}
\usepackage{multirow}
\usepackage{hhline}
\usepackage{bbm}
\usepackage{float}
\usepackage{epsfig}
\usepackage{graphics}  
\usepackage{graphicx}  
\usepackage{dcolumn}   
\usepackage{bm}        
\usepackage{amssymb}   
\usepackage{amsmath}   
\usepackage{mathrsfs}  
\usepackage{booktabs} 
\usepackage{siunitx} 
\usepackage{array} 
\usepackage{caption} 
\usepackage{amsfonts}
\usepackage{cancel}
\usepackage[percent]{overpic}
\usepackage{natbib}
\usepackage{hyperref}
\usepackage[usenames]{color}
\usepackage{CJK}

\usepackage{tikz,xcolor,hyperref}

\definecolor{lime}{HTML}{A6CE39}
\DeclareRobustCommand{\orcidicon}{
	\begin{tikzpicture}
	\draw[lime, fill=lime] (0,0) 
	circle [radius=0.16] 
	node[white] {{\fontfamily{qag}\selectfont \tiny ID}};
	\draw[white, fill=white] (-0.0625,0.095) 
	circle [radius=0.007];
	\end{tikzpicture}
	\hspace{-2mm}
}
\foreach \x in {A, ..., Z}{%
	\expandafter\xdef\csname orcid\x\endcsname{\noexpand\href{https://orcid.org/\csname orcidauthor\x\endcsname}{\noexpand\orcidicon}}
}
\foreach \x in {A, ..., Z}{%
	\expandafter\xdef\csname orcid\x\endcsname{\noexpand\href{https://orcid.org/\csname orcidauthor\x\endcsname}{\noexpand\orcidicon}}
}

\begin{document}

\title{Minimal Wigner-$SU(4)$ Interaction in Microscopic Cluster Models for $\alpha$-Conjugate Nuclei
}

\author{Guo-Ping Li}
\affiliation{Key Laboratory of Nuclear Physics and Ion-beam Application (MOE), Institute of Modern Physics, Fudan University, Shanghai 200433, China}
\author{Su-Yu Zhou}
\affiliation{Key Laboratory of Nuclear Physics and Ion-beam Application (MOE), Institute of Modern Physics, Fudan University, Shanghai 200433, China}
\author{Dong Bai}
\affiliation{College of Mechanics and Engineering Science, Hohai University, Nanjing, 211100, China}
\author{Bo Zhou\orcidD{}}
 \email{zhou\_bo@fudan.edu.cn}
 \affiliation{Key Laboratory of Nuclear Physics and Ion-beam Application (MOE), Institute of Modern Physics, Fudan University, Shanghai 200433, China}
 \affiliation{Shanghai Research Center for Theoretical Nuclear Physics, NSFC and Fudan University, Shanghai 200438, China}
 \author{Yu-Gang~Ma\orcidE{}}
 \email{mayugang@fudan.edu.cn}
 \affiliation{Key Laboratory of Nuclear Physics and Ion-beam Application (MOE), Institute of Modern Physics, Fudan University, Shanghai 200433, China}
 \affiliation{Shanghai Research Center for Theoretical Nuclear Physics, NSFC and Fudan University, Shanghai 200438, China}
 \affiliation{School of Physics, East China Normal University, Shanghai 200241, China}

\begin{abstract}
We present a minimalist, symmetry-guided interaction for microscopic cluster models based on Wigner-$SU(4)$ symmetry. Retaining only an $SU(4)$-invariant two-body attraction and a local three-body repulsion, this framework is implemented via the generator coordinate method (GCM) to describe $\alpha$--$\alpha$ scattering phase shifts, the low-lying spectrum and transition properties of $^{12}\mathrm{C}$, and the cluster spectrum of $^{16}\mathrm{O}$. We show that the long-standing structural tension between the $^{12}\mathrm{C}$ and $^{16}\mathrm{O}$ ground states can be mitigated within this restricted $SU(4)$ operator space without introducing additional phenomenological complexity. These results indicate that Wigner-$SU(4)$ symmetry provides an effective organizing principle for $N\alpha$ clustering, offering a more fundamental baseline for understanding complex cluster structures.
\end{abstract}
\maketitle

\emph{Introduction}{\bf ---}
The microscopic description of clustering in light nuclei, particularly within $N\alpha$ systems, is a long-standing yet actively evolving frontier in nuclear many-body physics \cite{vonoertzen:2006, Freer:2014qoa, Freer:2017gip}. It encompasses a rich array of phenomena, providing essential insights into stellar nucleosynthesis \cite{Sun:2022xjr}, high-energy heavy-ion collisions \cite{wanbinghe:2014prl,Br2014,Zhang2017,MaZhang:2020,Li2026}, and exotic states analogous to the Bose-Einstein condensate (BEC) \cite{Ropke:1998qs,Zhou:2023vgv}. As illustrated by the Ikeda diagram \cite{ikeda:1972}, developed $N\alpha$ cluster structures emerge near the breakup thresholds. Microscopic cluster models, notably the GCM, serve as a versatile framework for investigating these $\alpha$-cluster structures \cite{zhou_nonlocalized_2019c, Liu2025}. 

However, a persistent difficulty remains in the microscopic description of $N\alpha$ clustering. Originally proposed in 1965, the Volkov interaction \cite{VOLKOV196533} remains highly popular in nuclear cluster models for describing $N\alpha$ systems~\cite{Freer:2017gip}, as it successfully reproduces the energy spectrum and related observables of $^{12}\mathrm{C}$~\cite{uegaki1977,Zhou:2020nonlocalized}. Yet this effective nucleon-nucleon interaction completely fails to reproduce the $4\alpha$ cluster structure of $^{16}\mathrm{O}$. To address this deficiency, various improved interactions have been developed over the years, including finite-range three-body effective forces \cite{Tohsaki:1994zz,enyo:2004}, extended Volkov-type interactions with additional Bartlett and Heisenberg exchange terms \cite{Dufour:2005boe}, and realistic-interaction-based potentials \cite{Itagaki:2020pib}. Despite these advancements, achieving a unified description of the $3\alpha$ and $4\alpha$ systems remains a major obstacle. Specifically, these potentials struggle to simultaneously reconcile the highly developed, dilute $3\alpha$ cluster configurations in $^{12}\mathrm{C}$ with the compact, shell-model-like $4\alpha$ ground state of $^{16}\mathrm{O}$ \cite{Irvine01111971,mitchell:1964, Itagaki:1995, Itagaki:2016bxb, Itagaki:2020pib}. Employing more microscopic or even bare nuclear interactions might appear to be a preferable alternative \cite{Hammer:2019poc,Epelbaum:2020fip,MACHLEIDT:2024}. Nevertheless, the inherent complexity of such chiral interactions not only imposes prohibitive computational demands, but also makes them exceedingly difficult to adapt into the conventional cluster model framework, which remains a prominent challenge today \cite{Fukui:2020ylj}.

A possible way to understand this structural duality lies in the fundamental symmetries of nuclear forces. In recent years, while chiral effective field theory (ChEFT) has seen widespread success in nuclear structure calculations \cite{Hammer:2019poc,Epelbaum:2020fip,MACHLEIDT:2024,wu2026search}, a deeper structural insight emerges from the large-$N_c$ limit of quantum chromodynamics (QCD). In this limit, the leading-order (LO) nucleon-nucleon interaction naturally respects Wigner-$SU(4)$ symmetry \cite{Wigner:1936dx, Kaplan:1995yg, Kaplan:1996rk}. Recent advancements in ab initio nuclear lattice effective field theory (NLEFT) have shown that $SU(4)$-symmetric interactions provide an efficient starting point for describing spatial correlations and clustering in light nuclei \cite{Lu:2018bat, niu:2025spf, Shen:2022bak, Shen:2024qzi}. Alternatively, this emergent Wigner-$SU(4)$ symmetry can also be viewed as a consequence of the interplay between the unitary and chiral limits \cite{Lyu:2025yhz}.

In this paper, we construct a minimal Wigner-$SU(4)$-symmetric effective interaction tailored to microscopic cluster models. Implemented in the GCM framework, this Hamiltonian captures the main low-lying structures of $\alpha$-conjugate nuclei. By comparing two representative parametrizations within the same restricted $SU(4)$-symmetric operator space, we show that this minimal structure helps reduce the persistent structural tension between $^{12}\mathrm{C}$ and $^{16}\mathrm{O}$ without introducing additional phenomenological operators. This supports Wigner-$SU(4)$ symmetry as a practical leading baseline for microscopic studies of $N\alpha$ systems.

\emph{Method}{\bf ---}
We employ the GCM with parity- and angular-momentum-projected Brink-Bloch basis states~\cite{brink1966alpha}. For an $\alpha$-conjugate system with $4n$ nucleons, the intrinsic wave function is constructed from $n$ localized $\alpha$ clusters and is fully antisymmetrized:
\begin{equation}
\label{eq:GCMWaveFunction}
\Psi_{n\alpha}^{\mathrm{B}}(\{\boldsymbol{R}\}) = \mathcal{A} \prod_{i=1}^{n} \left[ \prod_{m=1}^{4} \phi_{m}(\boldsymbol{r}_{k}; \boldsymbol{R}_{i}) \right] = \mathcal{A} \prod_{i=1}^{n} \psi_{\alpha}(\boldsymbol{R}_{i}).
\end{equation}
Here, $\{\boldsymbol{R}\} \equiv \{\boldsymbol{R}_1, \cdots, \boldsymbol{R}_n\}$ denotes the set of generator coordinates, where each vector $\boldsymbol{R}_i$ specifies the center-of-mass position of the $i$-th $\alpha$ cluster in coordinate space. Collectively, the set $\{\boldsymbol{R}\}$ defines the instantaneous spatial geometry and the multi-dimensional configuration space of the $n\alpha$ system. The index $m$ labels the four distinct spin-isospin states, and $k = 4(i-1) + m$. The single-nucleon wave packet is described by a standard Gaussian, $\phi_m(\boldsymbol{r}_k; \boldsymbol{R}_i) = (1/\pi b^2)^{3/4} \exp[-(\boldsymbol{r}_k - \boldsymbol{R}_i)^2/(2b^2)] \chi_{\sigma\tau}^{(k)}$, where the harmonic-oscillator size parameter $b$ is set to $1.46\ \mathrm{fm}$, a conventional value used in microscopic cluster models. This intrinsic construction naturally enforces the internal state of each $\alpha$-cluster to be a scalar $SU(4)$ singlet, inherently aligning with the underlying Wigner-$SU(4)$ symmetry of our effective Hamiltonian.

To restore the relevant symmetries and determine the physical spectrum, the total wave function $\Psi^{J^\pi}_{\mathrm{GCM}}$ with definite angular momentum $J$ and parity $\pi$ is constructed by superposing the projected intrinsic states over various geometric configurations $\{\boldsymbol{R}\}$. The weight coefficients and energy eigenvalues are obtained by solving the Hill-Wheeler-Griffin equation \cite{Hill:1952jb}.

The microscopic Hamiltonian is expressed as:
\begin{equation}
	\label{eq:Hamiltonian}
    H=\sum_{i}T_{i}-T_{\mathrm{c.m.}}+\sum_{i< j} V^{\mathrm{C}}_{ij}+V^{\mathrm{N}}_{SU(4)},
\end{equation}
where $T_i$ and $T_{\mathrm{c.m.}}$ are the single-nucleon and center-of-mass kinetic energy operators, respectively. $V^{\mathrm{C}}_{ij}$ and $V^{\mathrm{N}}_{SU(4)}$ denote the Coulomb and the Wigner-$SU(4)$ invariant nuclear interactions. 

Rather than attempting a full order-by-order ChEFT construction, we use Wigner-$SU(4)$ symmetry to restrict the operator content and adopt a practical local effective interaction compatible with Brink-GCM calculations, comprising two- and three-body central terms:
\begin{equation}
\begin{aligned}
    V^{\mathrm{N}}_{SU(4)} =& C \sum_{i < j} \exp\left( -\frac{\Lambda^2 r_{i j}^{2}}{4\hbar^2 c^2} \right) \\
    & \quad + D \sum_{i< j < k} \exp\left( -\frac{\Lambda^2 (r_{i j}^{2}+r_{ik}^{2}+r_{jk}^{2})}{4\hbar^2 c^2} \right),
\end{aligned}
\end{equation}
where $C$ and $D$ are the low-energy constants (LECs). The resolution scale $\Lambda$ governs the range of the Gaussian regulator, effectively treating the interaction as a locally regulated contact force. In contrast to conventional effective nucleon-nucleon interactions (e.g., the Volkov force \cite{VOLKOV196533}), this minimal formulation omits all Majorana, Bartlett, and Heisenberg exchange operators, relying solely on the scalar $SU(4)$ singlet configuration to drive dynamics. The three-body term is physically indispensable; motivated by large-$N_c$ analyses in which leading local three-nucleon force structures provide dominant contributions~\cite{Phillips:2013rsa,Epelbaum:2014sea}, it not only ensures proper density saturation to avert unphysical collapse but also helps govern the inter-cluster separation. 

\begin{table}[htbp]
\captionsetup{singlelinecheck=false, justification=raggedright} 
\caption{Determined values of the parameters $C$ and $D$. SU4A is anchored to the ground state of $^4\mathrm{He}$ and the ground state and Hoyle state of $^{12}\mathrm{C}$; SU4B simultaneously includes the ground state of $^{12}\mathrm{C}$ and $^{16}\mathrm{O}$ in the fit. All parameters are in MeV.}
\centering
\renewcommand{\arraystretch}{1.1}
\begin{tabular}{c c c c}
\hline\hline
\rule{0pt}{3ex}
&$\Lambda$ & $C$  & $D$ \\
\hline
SU4A & 500 & -316.2 & 1187.4 \\
SU4B & 400 & -180.3 & 415.5 \\
\hline\hline
\end{tabular}%
\label{tab:eft_parameters}
\end{table}

\emph{Results and discussion}{\bf ---}
With only two parameters, $C$ and $D$, plus the resolution scale $\Lambda$, the interaction is determined by two strategies, summarized in Table~\ref{tab:eft_parameters}. The first strategy, denoted SU4A, anchors the parameters to the ground-state energies of $^4\mathrm{He}$ and $^{12}\mathrm{C}$ at an optimized resolution scale of $\Lambda = 500$ MeV (as detailed in the Supplemental Material \cite{supplement}). The second strategy, denoted SU4B, represents a more standard and comprehensive approach for fixing effective interactions in microscopic cluster models. By simultaneously considering both the $3\alpha$ and $4\alpha$ systems, $C$ and $D$ are reoptimized to balance the overall ground-state energies of $^{12}\mathrm{C}$ and $^{16}\mathrm{O}$. This dual approach decouples the physical capabilities of the minimal $SU(4)$ operator structure from the limitations of fitting parameters to a single mass region.

\begin{figure}[htbp]
\centering
\includegraphics[width=0.4\textwidth]{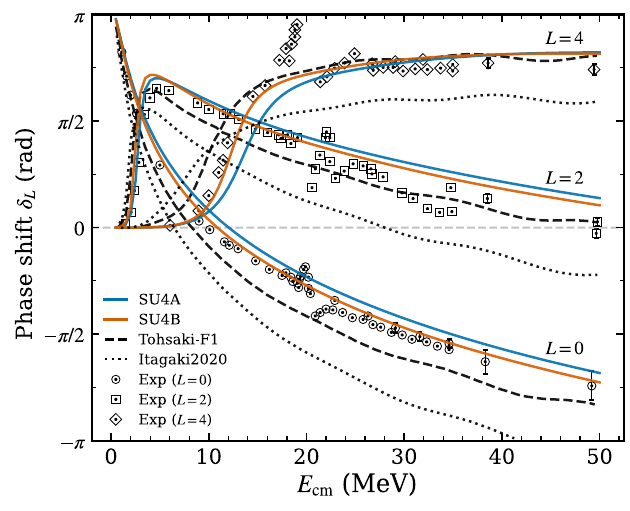} %
\captionsetup{singlelinecheck=false, justification=raggedright}
\caption{The scattering phase shifts of the $\alpha + \alpha$ system with SU4A, SU4B, Tohsaki-F1~\cite{Tohsaki:1994zz} and the results of Itagaki \textit{et al.}~\cite{Itagaki:2020pib}. The experimental phase shifts $\delta_L$ are taken from Ref.~\cite{Fukui:2020ylj}.}
\label{fig:aaphaseshift}
\end{figure}

To validate the proposed interaction in open-channel dynamics, we calculate the $\alpha$--$\alpha$ scattering phase shifts using the Kohn-Hulth\'{e}n variational method within the GCM framework \cite{Mito1976,Kamimura:1977okl} (see Fig.~\ref{fig:aaphaseshift}). Both SU4A and SU4B reproduce the main trends of the experimental $\alpha$--$\alpha$ phase shifts, with SU4B giving a particularly close description over the energy range shown. This comparison indicates that the essential $\alpha$-$\alpha$ dynamics can be accommodated within the same minimal $SU(4)$-symmetric operator structure under different calibration strategies.

\begin{table}[htbp]
\centering
\captionsetup{singlelinecheck=false, justification=raggedright}
\caption{Energies, radii and transition rates of $^{12}\mathrm{C}$. The energies are measured from the $3\alpha$ threshold. Here, only the central energy values of the NLEFT results are considered in our table. Here, $r$ denotes the matter radius. For $r_c(0_1^+)$, the charge radius of the proton $r^p_E = 0.840\ \mathrm{fm}$ \cite{Lin:2021xrc} is added in quadrature. All energies are in MeV, radii in fm, $M(E0)$ in $e\ \mathrm{fm}^2$, and $B(E2)$ in $e^2\ \mathrm{fm}^4$.} 
\resizebox{0.5\textwidth}{!}{%
\begin{tabular}{lcccccc@{}}
\hline\hline
\rule{0pt}{3ex}
         & Exp. & GCM-SU4A & GCM-SU4B & NLEFT \cite{Shen:2022bak} & REM \cite{Imai:2018lww} & THSR \cite{Funaki:2014tda,Funaki:2016atc} \\
\hline
$E(0_1^+)$  & -7.3 & -7.3 & -6.5 & -6.7 & -7.6 & -7.5 \\
$E(0_2^+)$ & 0.4 & 0.4 & 0.0 & -0.8 & 0.3 & 0.2\\
$E(0_3^+)$  & 3.0 (3) & 3.1 & 2.0 & 2.2 & 2.8 & 3.9 \\
$E(2_1^+)$ & -2.8 & -3.2 & -3.0 & -2.0 & -5.1 & -4.8 \\
\hline
$r_c(0_1^+)$ & 2.47 (2) & 2.53 & 2.63 & 2.54 (1) & 2.54 & 2.54 \\
$r(0_2^+)$ & - & 3.81 & 3.85 & 3.45(2) & 3.7 & 3.7 \\
$r(0_3^+)$  & - & 4.89 & 4.95 & 3.47(1) & 4.6 & 4.7  \\
$r(2_1^+)$  & - & 2.39 & 2.50 & 2.42(1)(1) & 2.4 & 2.4 \\
\hline
$M(E0, 0_1^+ \rightarrow 0_2^+)$   & 5.4(2) & 5.9 & 6.9 & 4.8 (3) & 6.4 & 6.3 \\
$M(E0, 0_1^+ \rightarrow 0_3^+)$  & - & 3.0 & 2.8 & 0.4 (3) & 3.8 & 3.9\\
$M(E0, 0_2^+ \rightarrow 0_3^+)$ & - & 39.2 & 39.1 & 7.4 (4) & 28 & 34\\
\hline
$B(E2, 2_1^+ \rightarrow 0_1^+)$   & 7.9(4) & 8.6 & 11.9 & 11.4(1)(4.3) & - & 9.5 \\
$B(E2, 2_1^+ \rightarrow 0_2^+)$  & 2.6(4) & 2.0 & 3.0 & 2.4(2)(7) & - & 1.0 \\
\hline\hline
\end{tabular}}
\label{tab:c12_properties}
\end{table}

\begin{figure*}[t!]
\centering
\includegraphics[width=1.0\textwidth]{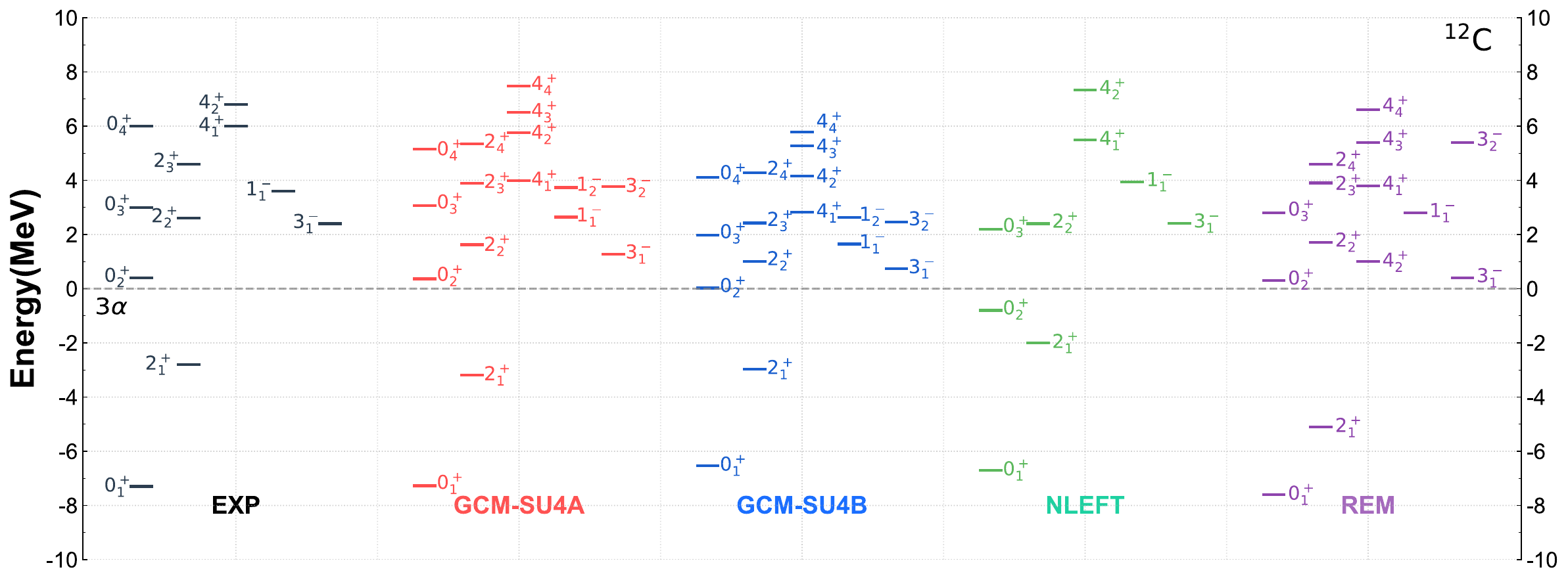} %
\captionsetup{singlelinecheck=false, justification=raggedright}
\caption{Energy levels of $^{12}\mathrm{C}$, measured from the $3\alpha$ threshold, are presented separately according to their spin-parity $J^{\pi}$. The GCM results utilizing the SU4A and SU4B interactions, NLEFT \cite{Shen:2022bak}, and REM \cite{Imai:2018lww} are compared with experimental data \cite{Kelley:2017qgh}.}
\label{fig:12Cenergy}
\end{figure*}

Using the two interactions, we calculate the low-lying spectra and transition rates of $^{12}\mathrm{C}$ via the GCM superposition of 695 $3\alpha$ Brink configurations. The results, summarized in Fig. \ref{fig:12Cenergy} and Table \ref{tab:c12_properties}, are compared with those of \textit{ab initio} NLEFT \cite{Shen:2022bak}, REM \cite{Imai:2018lww}, THSR \cite{Funaki:2014tda,Funaki:2016atc}, and experiment~\cite{Ajzenberg-Selove:1990fsm, Angeli:2013epw, Kelley:2017qgh}. As illustrated in Fig.~\ref{fig:12Cenergy}, our method provides a good description of the $2_1^+$ state energy compared to conventional phenomenological interactions. It is well known that traditional cluster model interactions underestimate this $2_1^+$ excitation energy — a chronic issue that typically requires advanced frameworks like antisymmetrized molecular dynamics (AMD) to partially resolve. Furthermore, the observables derived from the wave functions are in agreement with experimental data; for example, the charge radius of the ground state and the $E0$ transition matrix element from the structurally sensitive Hoyle ($0_2^+$) state are well reproduced. Overall, the two parametrizations give a reasonable description of the available $^{12}\mathrm{C}$ observables and remain broadly consistent with other microscopic cluster calculations. However, while a precise description of $^{12}\mathrm{C}$ is encouraging, it is not entirely surprising for a well-calibrated model; the true, highly anticipated test of our framework lies in its ability to capture the complex $4\alpha$ dynamics of $^{16}\mathrm{O}$.

\begin{figure*}[!t]
\centering
\includegraphics[width=1.0\textwidth]{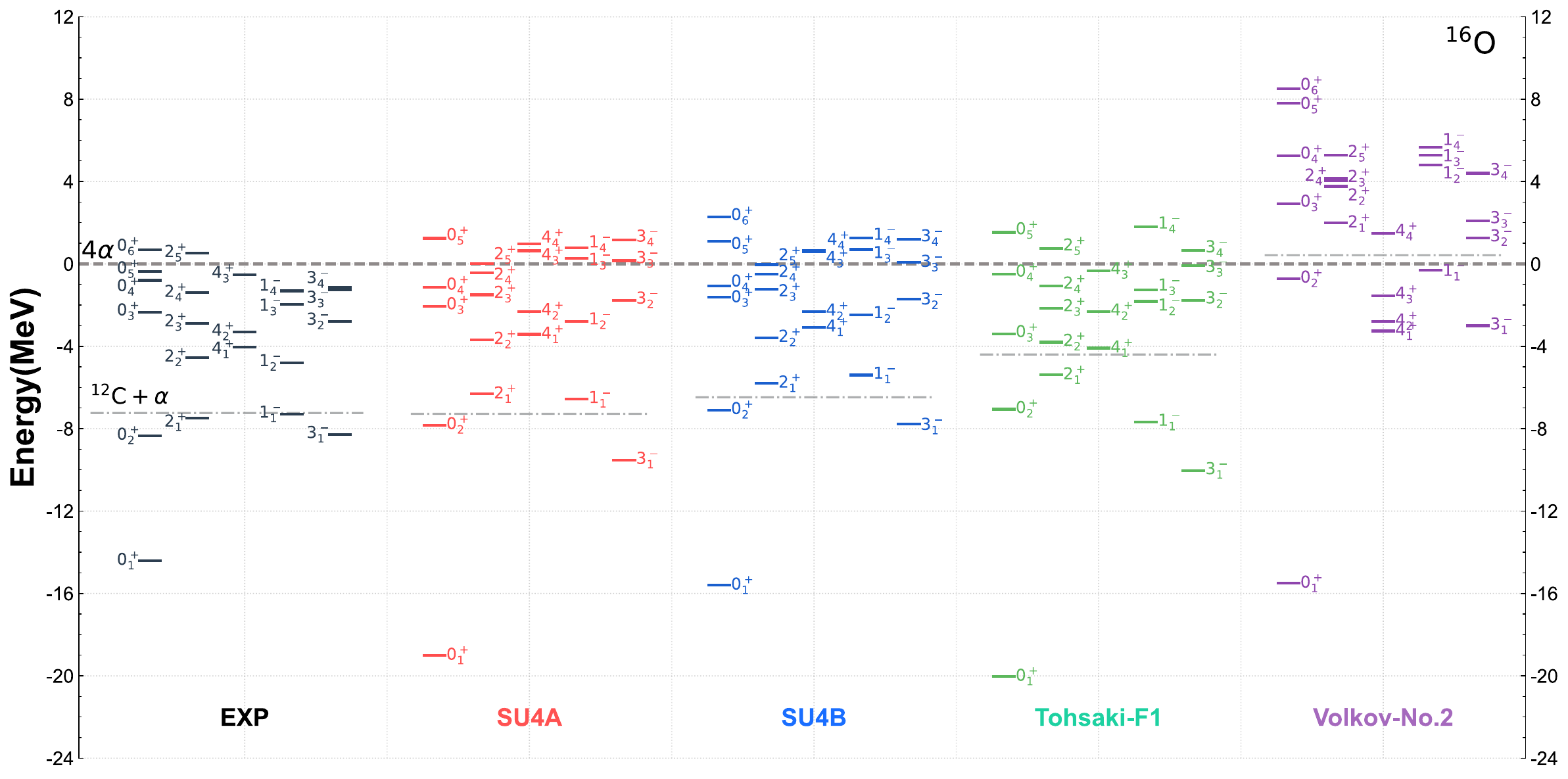} %
\captionsetup{singlelinecheck=false, justification=raggedright}
\caption{Energy levels for $^{16}\mathrm{O}$ from experimental data \cite{Ajzenberg-Selove:1982fgy, Wakasa:2006nt}, GCM results with SU4A, SU4B, Tohsaki-F1 force \cite{syzhou:2025}, and Volkov-No.2 potential (implemented within our GCM framework using parameters from Ref. \cite{Dufour:2014rta} and verified to reproduce its reported ground state). The gray dashed and dash-dotted lines represent the thresholds of $4\alpha$ and $^{12}\mathrm{C}+\alpha$, respectively.}
\label{fig:16Oenergy}
\end{figure*}

A major obstacle for microscopic cluster models in accurately describing the $4\alpha$ cluster structure of $^{16}\mathrm{O}$ stems directly from the deficiencies of the underlying interactions. The $^{16}\mathrm{O}$ calculation superposes 795 Brink-GCM basis states, including $4\alpha$ and $^{12}\mathrm{C}+\alpha$ configurations. As shown in Fig.~\ref{fig:16Oenergy}, traditional interactions like the Volkov-No.2 potential fail to reproduce the correct threshold behavior, placing the $^{12}\mathrm{C}+\alpha$ threshold far too deeply (deviating by more than 7.7 MeV relative to the experimental placement). Even the Tohsaki-F1 force suffers from a severe overbinding of the $0_1^+$ ground state, overshooting the experimental level by roughly 5.6 MeV. Relative to the $4\alpha$ threshold, SU4A gives $E(0_1^+)\simeq -19.1~\mathrm{MeV}$, overbinding the ground state by about $4.6~\mathrm{MeV}$ relative to the experimental value. SU4B moves this state to $E(0_1^+)\simeq -15.6~\mathrm{MeV}$, leaving a residual deviation of about $1.2~\mathrm{MeV}$. At the same time, SU4B largely reproduces the placement of the $^{12}\mathrm{C}+\alpha$ threshold and gives a reasonable description of the low-lying cluster spectrum. Considering its description of the $\alpha$--$\alpha$ scattering phase shifts and the $^{12}\mathrm{C}$ properties, SU4B substantially overcomes the fundamental limitations of both traditional and recently improved interactions. These results may provide potentially useful microscopic input for future studies of the astrophysical $^{12}\mathrm{C}(\alpha,\gamma)^{16}\mathrm{O}$ reaction~\cite{deboer:2017rmp}, $^{12}\mathrm{C}$ and $^{16}\mathrm{O}$ fusion reactions~\cite{cljiang:2007fusion}, high-energy heavy-ion collisions for studying light nuclei~\cite{Li2020, ALICE:2026geometryOO,CMS:2026OONeNeFlow}, and the search for new highly excited cluster states above the $4\alpha$ threshold~\cite{chen_new_2023-1}.

As established in the seminal works of Wigner \cite{Wigner:1936dx} and Volkov \cite{VOLKOV196533}, the interplay between $SU(4)$ symmetry and the Pauli principle dictates the maximum spatial symmetry in the ground states of $\alpha$-conjugate nuclei. Our Hamiltonian produces clustered structures through two mechanisms: an attractive two-body term that leads to spatially correlated $N\alpha$ clusters~\cite{lee:2007sca}, and a local three-body repulsion that ensures saturation. Low-energy $\alpha$-clustering is dominated by such symmetric configurations. By excluding explicit $SU(4)$-breaking exchange and non-central terms, the present central interaction provides a clean baseline for examining the $SU(4)$-symmetric contribution to low-energy $N\alpha$ dynamics.

As a further test of transferability, we apply the present interaction to the open-shell $5\alpha$ system $^{20}\mathrm{Ne}$. In the GCM-Brink calculations, the SU4A parametrization yields a ground-state energy $E_{\mathrm{gs}} = -161.80$ MeV and a charge radius $r_{\mathrm{c}} = 2.93$ fm, which show reasonable agreement with the experimental values of $-160.65$ MeV and 3.0055(21) fm. Conversely, SU4B produces an underbound ground state ($E_{\mathrm{gs}} = -157.56$ MeV, $r_{\mathrm{c}} = 3.08$ fm) and a noticeably compressed low-lying spectrum. Compared to traditional effective interactions, such as the Volkov force—which fails to reproduce both the binding energy and the charge radius of $^{20}\mathrm{Ne}$ even with adjusted Majorana parameters — the SU4A result demonstrates the potential of the minimal $SU(4)$ framework. The observed discrepancies, particularly for SU4B, suggest that while this minimal interaction captures the bulk features of $N\alpha$ dynamics, its precision in heavier open-shell systems may be sensitive to the globally fixed harmonic-oscillator size parameter $b = 1.46$ fm. A dynamic adjustment of the cluster size or an inclusion of state-dependent effects might be required to further refine the description of such systems. The calculated low-lying energy levels of $^{20}\mathrm{Ne}$ are provided in the Supplemental Material~\cite{supplement}. 

\emph{Summary}{\bf ---}
In summary, we have developed a Wigner-$SU(4)$-symmetric interaction with a two-body attraction and a three-body repulsion, and applied it to $\alpha$-conjugate nuclei from $^{4}\mathrm{He}$ to $^{20}\mathrm{Ne}$. By examining two representative parametrizations of the same interaction form, SU4A and SU4B, we show that this compact Hamiltonian describes key features of $\alpha$--$\alpha$ scattering, the low-lying structure of $^{12}\mathrm{C}$, and the cluster spectrum of $^{16}\mathrm{O}$. The comparison between SU4A and SU4B indicates that the long-standing $^{12}\mathrm{C}$--$^{16}\mathrm{O}$ structural tension can be reduced without introducing additional $SU(4)$-breaking operators. Calculations up to $^{20}\mathrm{Ne}$ further indicate that this symmetry-guided interaction captures important bulk features of $N\alpha$ dynamics. Future work should test how far this interaction can be extended to reaction observables and heavier $\alpha$-conjugate systems, and a broader range of nuclear systems, including self-conjugate odd-odd nuclei and isotopic chains.

\begin{acknowledgements}
We are grateful for discussions with Prof. Evgeny Epelbaum and Prof. Bingwei Long. This work was supported in part by the National Key R\&D Program of China under Grant Nos. 2023YFA1606701, 2024YFA1610802, and 2018YFE0104600; the National Natural Science Foundation of China under Grant Nos. 12275054 and 12547102; the Guangdong Major Project of Basic and Applied Basic Research under Grant No. 2020B0301030008; the Shanghai Pilot Program for Basic Research - Fudan University under Grant No. 21TQ1400100 (22TQ006); and the STCSM under Grant No. 23590780100.
\end{acknowledgements}

%


\begin{thebibliography}{65}%
\makeatletter
\providecommand \@ifxundefined [1]{%
 \@ifx{#1\undefined}
}%
\providecommand \@ifnum [1]{%
 \ifnum #1\expandafter \@firstoftwo
 \else \expandafter \@secondoftwo
 \fi
}%
\providecommand \@ifx [1]{%
 \ifx #1\expandafter \@firstoftwo
 \else \expandafter \@secondoftwo
 \fi
}%
\providecommand \natexlab [1]{#1}%
\providecommand \enquote  [1]{``#1''}%
\providecommand \bibnamefont  [1]{#1}%
\providecommand \bibfnamefont [1]{#1}%
\providecommand \citenamefont [1]{#1}%
\providecommand \href@noop [0]{\@secondoftwo}%
\providecommand \href [0]{\begingroup \@sanitize@url \@href}%
\providecommand \@href[1]{\@@startlink{#1}\@@href}%
\providecommand \@@href[1]{\endgroup#1\@@endlink}%
\providecommand \@sanitize@url [0]{\catcode `\\12\catcode `\$12\catcode `\&12\catcode `\#12\catcode `\^12\catcode `\_12\catcode `\%12\relax}%
\providecommand \@@startlink[1]{}%
\providecommand \@@endlink[0]{}%
\providecommand \url  [0]{\begingroup\@sanitize@url \@url }%
\providecommand \@url [1]{\endgroup\@href {#1}{\urlprefix }}%
\providecommand \urlprefix  [0]{URL }%
\providecommand \Eprint [0]{\href }%
\providecommand \doibase [0]{https://doi.org/}%
\providecommand \selectlanguage [0]{\@gobble}%
\providecommand \bibinfo  [0]{\@secondoftwo}%
\providecommand \bibfield  [0]{\@secondoftwo}%
\providecommand \translation [1]{[#1]}%
\providecommand \BibitemOpen [0]{}%
\providecommand \bibitemStop [0]{}%
\providecommand \bibitemNoStop [0]{.\EOS\space}%
\providecommand \EOS [0]{\spacefactor3000\relax}%
\providecommand \BibitemShut  [1]{\csname bibitem#1\endcsname}%
\let\auto@bib@innerbib\@empty
\bibitem [{\citenamefont {{von Oertzen}}\ \emph {et~al.}(2006)\citenamefont {{von Oertzen}}, \citenamefont {Freer},\ and\ \citenamefont {{Kanada-En'yo}}}]{vonoertzen:2006}%
  \BibitemOpen
  \bibfield  {author} {\bibinfo {author} {\bibfnamefont {W.}~\bibnamefont {{von Oertzen}}}, \bibinfo {author} {\bibfnamefont {M.}~\bibnamefont {Freer}},\ and\ \bibinfo {author} {\bibfnamefont {Y.}~\bibnamefont {{Kanada-En'yo}}},\ }\href {https://doi.org/10.1016/j.physrep.2006.07.001} {\bibfield  {journal} {\bibinfo  {journal} {Phys. Rep.}\ }\textbf {\bibinfo {volume} {432}},\ \bibinfo {pages} {43} (\bibinfo {year} {2006})}\BibitemShut {NoStop}%
\bibitem [{\citenamefont {Freer}\ and\ \citenamefont {Fynbo}(2014)}]{Freer:2014qoa}%
  \BibitemOpen
  \bibfield  {author} {\bibinfo {author} {\bibfnamefont {M.}~\bibnamefont {Freer}}\ and\ \bibinfo {author} {\bibfnamefont {H.}~\bibnamefont {Fynbo}},\ }\href {https://doi.org/10.1016/j.ppnp.2014.06.001} {\bibfield  {journal} {\bibinfo  {journal} {Prog. Part. Nucl. Phys.}\ }\textbf {\bibinfo {volume} {78}},\ \bibinfo {pages} {1} (\bibinfo {year} {2014})}\BibitemShut {NoStop}%
\bibitem [{\citenamefont {Freer}\ \emph {et~al.}(2018)\citenamefont {Freer}, \citenamefont {Horiuchi}, \citenamefont {{Kanada-En'yo}}, \citenamefont {Lee},\ and\ \citenamefont {Mei{\ss}ner}}]{Freer:2017gip}%
  \BibitemOpen
  \bibfield  {author} {\bibinfo {author} {\bibfnamefont {M.}~\bibnamefont {Freer}}, \bibinfo {author} {\bibfnamefont {H.}~\bibnamefont {Horiuchi}}, \bibinfo {author} {\bibfnamefont {Y.}~\bibnamefont {{Kanada-En'yo}}}, \bibinfo {author} {\bibfnamefont {D.}~\bibnamefont {Lee}},\ and\ \bibinfo {author} {\bibfnamefont {U.-G.}\ \bibnamefont {Mei{\ss}ner}},\ }\href {https://doi.org/10.1103/RevModPhys.90.035004} {\bibfield  {journal} {\bibinfo  {journal} {Rev. Mod. Phys.}\ }\textbf {\bibinfo {volume} {90}},\ \bibinfo {pages} {035004} (\bibinfo {year} {2018})}\BibitemShut {NoStop}%
\bibitem [{\citenamefont {Sun}\ \emph {et~al.}(2024)\citenamefont {Sun}, \citenamefont {Wang}, \citenamefont {Ko}, \citenamefont {Ma},\ and\ \citenamefont {Shen}}]{Sun:2022xjr}%
  \BibitemOpen
  \bibfield  {author} {\bibinfo {author} {\bibfnamefont {K.-J.}\ \bibnamefont {Sun}}, \bibinfo {author} {\bibfnamefont {R.}~\bibnamefont {Wang}}, \bibinfo {author} {\bibfnamefont {C.~M.}\ \bibnamefont {Ko}}, \bibinfo {author} {\bibfnamefont {Y.-G.}\ \bibnamefont {Ma}},\ and\ \bibinfo {author} {\bibfnamefont {C.}~\bibnamefont {Shen}},\ }\href {https://doi.org/10.1038/s41467-024-45474-x} {\bibfield  {journal} {\bibinfo  {journal} {Nat. Commun.}\ }\textbf {\bibinfo {volume} {15}},\ \bibinfo {pages} {1074} (\bibinfo {year} {2024})}\BibitemShut {NoStop}%
\bibitem [{\citenamefont {He}\ \emph {et~al.}(2014)\citenamefont {He}, \citenamefont {Ma}, \citenamefont {Cao}, \citenamefont {Cai},\ and\ \citenamefont {Zhang}}]{wanbinghe:2014prl}%
  \BibitemOpen
  \bibfield  {author} {\bibinfo {author} {\bibfnamefont {W.~B.}\ \bibnamefont {He}}, \bibinfo {author} {\bibfnamefont {Y.~G.}\ \bibnamefont {Ma}}, \bibinfo {author} {\bibfnamefont {X.~G.}\ \bibnamefont {Cao}}, \bibinfo {author} {\bibfnamefont {X.~Z.}\ \bibnamefont {Cai}},\ and\ \bibinfo {author} {\bibfnamefont {G.~Q.}\ \bibnamefont {Zhang}},\ }\href {https://doi.org/10.1103/PhysRevLett.113.032506} {\bibfield  {journal} {\bibinfo  {journal} {Phys. Rev. Lett.}\ }\textbf {\bibinfo {volume} {113}},\ \bibinfo {pages} {032506} (\bibinfo {year} {2014})}\BibitemShut {NoStop}%
\bibitem [{\citenamefont {Broniowski}\ and\ \citenamefont {Ruiz~Arriola}(2014)}]{Br2014}%
  \BibitemOpen
  \bibfield  {author} {\bibinfo {author} {\bibfnamefont {W.}~\bibnamefont {Broniowski}}\ and\ \bibinfo {author} {\bibfnamefont {E.}~\bibnamefont {Ruiz~Arriola}},\ }\href {https://doi.org/10.1103/PhysRevLett.112.112501} {\bibfield  {journal} {\bibinfo  {journal} {Phys. Rev. Lett.}\ }\textbf {\bibinfo {volume} {112}},\ \bibinfo {pages} {112501} (\bibinfo {year} {2014})}\BibitemShut {NoStop}%
\bibitem [{\citenamefont {Zhang}\ \emph {et~al.}(2017)\citenamefont {Zhang}, \citenamefont {Ma}, \citenamefont {Chen}, \citenamefont {He},\ and\ \citenamefont {Zhong}}]{Zhang2017}%
  \BibitemOpen
  \bibfield  {author} {\bibinfo {author} {\bibfnamefont {S.}~\bibnamefont {Zhang}}, \bibinfo {author} {\bibfnamefont {Y.~G.}\ \bibnamefont {Ma}}, \bibinfo {author} {\bibfnamefont {J.~H.}\ \bibnamefont {Chen}}, \bibinfo {author} {\bibfnamefont {W.~B.}\ \bibnamefont {He}},\ and\ \bibinfo {author} {\bibfnamefont {C.}~\bibnamefont {Zhong}},\ }\href {https://doi.org/10.1103/PhysRevC.95.064904} {\bibfield  {journal} {\bibinfo  {journal} {Phys. Rev. C}\ }\textbf {\bibinfo {volume} {95}},\ \bibinfo {pages} {064904} (\bibinfo {year} {2017})}\BibitemShut {NoStop}%
\bibitem [{\citenamefont {Ma}\ and\ \citenamefont {Zhang}(2020)}]{MaZhang:2020}%
  \BibitemOpen
  \bibfield  {author} {\bibinfo {author} {\bibfnamefont {Y.-G.}\ \bibnamefont {Ma}}\ and\ \bibinfo {author} {\bibfnamefont {S.}~\bibnamefont {Zhang}},\ }\bibinfo {title} {Influence of nuclear structure in relativistic heavy-ion collisions},\ in\ \href {https://doi.org/10.1007/978-981-15-8818-1_5-1} {\emph {\bibinfo {booktitle} {Handbook of Nuclear Physics}}},\ \bibinfo {editor} {edited by\ \bibinfo {editor} {\bibfnamefont {I.}~\bibnamefont {Tanihata}}, \bibinfo {editor} {\bibfnamefont {H.}~\bibnamefont {Toki}},\ and\ \bibinfo {editor} {\bibfnamefont {T.}~\bibnamefont {Kajino}}}\ (\bibinfo  {publisher} {Springer Nature Singapore},\ \bibinfo {address} {Singapore},\ \bibinfo {year} {2020})\ pp.\ \bibinfo {pages} {1--30}\BibitemShut {NoStop}%
\bibitem [{\citenamefont {Li}\ \emph {et~al.}(2026{\natexlab{a}})\citenamefont {Li}, \citenamefont {Zhou},\ and\ \citenamefont {Ma}}]{Li2026}%
  \BibitemOpen
  \bibfield  {author} {\bibinfo {author} {\bibfnamefont {P.}~\bibnamefont {Li}}, \bibinfo {author} {\bibfnamefont {B.}~\bibnamefont {Zhou}},\ and\ \bibinfo {author} {\bibfnamefont {G.~L.}\ \bibnamefont {Ma}},\ }\href {https://doi.org/10.1103/tffz-8q1m} {\bibfield  {journal} {\bibinfo  {journal} {Phys. Rev. Lett.}\ }\textbf {\bibinfo {volume} {136}},\ \bibinfo {pages} {082302} (\bibinfo {year} {2026}{\natexlab{a}})}\BibitemShut {NoStop}%
\bibitem [{\citenamefont {R{\"o}pke}\ \emph {et~al.}(1998)\citenamefont {R{\"o}pke}, \citenamefont {Schnell}, \citenamefont {Schuck},\ and\ \citenamefont {Nozi{\`e}res}}]{Ropke:1998qs}%
  \BibitemOpen
  \bibfield  {author} {\bibinfo {author} {\bibfnamefont {G.}~\bibnamefont {R{\"o}pke}}, \bibinfo {author} {\bibfnamefont {A.}~\bibnamefont {Schnell}}, \bibinfo {author} {\bibfnamefont {P.}~\bibnamefont {Schuck}},\ and\ \bibinfo {author} {\bibfnamefont {P.}~\bibnamefont {Nozi{\`e}res}},\ }\href {https://doi.org/10.1103/PhysRevLett.80.3177} {\bibfield  {journal} {\bibinfo  {journal} {Phys. Rev. Lett.}\ }\textbf {\bibinfo {volume} {80}},\ \bibinfo {pages} {3177} (\bibinfo {year} {1998})}\BibitemShut {NoStop}%
\bibitem [{\citenamefont {Zhou}\ \emph {et~al.}(2023)\citenamefont {Zhou}, \citenamefont {Funaki}, \citenamefont {Horiuchi}, \citenamefont {Ma}, \citenamefont {R{\"o}pke}, \citenamefont {Schuck}, \citenamefont {Tohsaki},\ and\ \citenamefont {Yamada}}]{Zhou:2023vgv}%
  \BibitemOpen
  \bibfield  {author} {\bibinfo {author} {\bibfnamefont {B.}~\bibnamefont {Zhou}}, \bibinfo {author} {\bibfnamefont {Y.}~\bibnamefont {Funaki}}, \bibinfo {author} {\bibfnamefont {H.}~\bibnamefont {Horiuchi}}, \bibinfo {author} {\bibfnamefont {Y.-G.}\ \bibnamefont {Ma}}, \bibinfo {author} {\bibfnamefont {G.}~\bibnamefont {R{\"o}pke}}, \bibinfo {author} {\bibfnamefont {P.}~\bibnamefont {Schuck}}, \bibinfo {author} {\bibfnamefont {A.}~\bibnamefont {Tohsaki}},\ and\ \bibinfo {author} {\bibfnamefont {T.}~\bibnamefont {Yamada}},\ }\href {https://doi.org/10.1038/s41467-023-43816-9} {\bibfield  {journal} {\bibinfo  {journal} {Nat. Commun.}\ }\textbf {\bibinfo {volume} {14}},\ \bibinfo {pages} {8206} (\bibinfo {year} {2023})}\BibitemShut {NoStop}%
\bibitem [{\citenamefont {Ikeda}\ \emph {et~al.}(1972)\citenamefont {Ikeda}, \citenamefont {Marumori}, \citenamefont {Tamagaki},\ and\ \citenamefont {Tanaka}}]{ikeda:1972}%
  \BibitemOpen
  \bibfield  {author} {\bibinfo {author} {\bibfnamefont {K.}~\bibnamefont {Ikeda}}, \bibinfo {author} {\bibfnamefont {T.}~\bibnamefont {Marumori}}, \bibinfo {author} {\bibfnamefont {R.}~\bibnamefont {Tamagaki}},\ and\ \bibinfo {author} {\bibfnamefont {H.}~\bibnamefont {Tanaka}},\ }\href {https://doi.org/10.1143/PTPS.52.1} {\bibfield  {journal} {\bibinfo  {journal} {Prog. Theor. Phys. Suppl.}\ }\textbf {\bibinfo {volume} {52}},\ \bibinfo {pages} {1} (\bibinfo {year} {1972})}\BibitemShut {NoStop}%
\bibitem [{\citenamefont {Zhou}\ \emph {et~al.}(2019)\citenamefont {Zhou}, \citenamefont {Funaki}, \citenamefont {Horiuchi},\ and\ \citenamefont {Tohsaki}}]{zhou_nonlocalized_2019c}%
  \BibitemOpen
  \bibfield  {author} {\bibinfo {author} {\bibfnamefont {B.}~\bibnamefont {Zhou}}, \bibinfo {author} {\bibfnamefont {Y.}~\bibnamefont {Funaki}}, \bibinfo {author} {\bibfnamefont {H.}~\bibnamefont {Horiuchi}},\ and\ \bibinfo {author} {\bibfnamefont {A.}~\bibnamefont {Tohsaki}},\ }\href {https://doi.org/10.1007/s11467-019-0917-0} {\bibfield  {journal} {\bibinfo  {journal} {Front. Phys.}\ }\textbf {\bibinfo {volume} {15}},\ \bibinfo {pages} {14401} (\bibinfo {year} {2019})}\BibitemShut {NoStop}%
\bibitem [{\citenamefont {Liu}\ \emph {et~al.}(2025)\citenamefont {Liu}, \citenamefont {Zhou},\ and\ \citenamefont {Ma}}]{Liu2025}%
  \BibitemOpen
  \bibfield  {author} {\bibinfo {author} {\bibfnamefont {Y.~F.}\ \bibnamefont {Liu}}, \bibinfo {author} {\bibfnamefont {B.}~\bibnamefont {Zhou}},\ and\ \bibinfo {author} {\bibfnamefont {Y.~G.}\ \bibnamefont {Ma}},\ }\href {https://doi.org/10.1007/s41365-025-01775-4} {\bibfield  {journal} {\bibinfo  {journal} {Nucl. Sci. Tech.}\ }\textbf {\bibinfo {volume} {36}},\ \bibinfo {pages} {196} (\bibinfo {year} {2025})}\BibitemShut {NoStop}%
\bibitem [{\citenamefont {Volkov}(1965)}]{VOLKOV196533}%
  \BibitemOpen
  \bibfield  {author} {\bibinfo {author} {\bibfnamefont {A.}~\bibnamefont {Volkov}},\ }\href {https://doi.org/https://doi.org/10.1016/0029-5582(65)90244-0} {\bibfield  {journal} {\bibinfo  {journal} {Nucl. Phys.}\ }\textbf {\bibinfo {volume} {74}},\ \bibinfo {pages} {33} (\bibinfo {year} {1965})}\BibitemShut {NoStop}%
\bibitem [{\citenamefont {Uegaki}\ \emph {et~al.}(1977)\citenamefont {Uegaki}, \citenamefont {Okabe}, \citenamefont {Abe},\ and\ \citenamefont {Tanaka}}]{uegaki1977}%
  \BibitemOpen
  \bibfield  {author} {\bibinfo {author} {\bibfnamefont {E.}~\bibnamefont {Uegaki}}, \bibinfo {author} {\bibfnamefont {S.}~\bibnamefont {Okabe}}, \bibinfo {author} {\bibfnamefont {Y.}~\bibnamefont {Abe}},\ and\ \bibinfo {author} {\bibfnamefont {H.}~\bibnamefont {Tanaka}},\ }\href {https://doi.org/10.1143/PTP.57.1262} {\bibfield  {journal} {\bibinfo  {journal} {Prog. Theor. Phys.}\ }\textbf {\bibinfo {volume} {57}},\ \bibinfo {pages} {1262} (\bibinfo {year} {1977})}\BibitemShut {NoStop}%
\bibitem [{\citenamefont {Zhou}\ \emph {et~al.}(2020)\citenamefont {Zhou}, \citenamefont {Funaki}, \citenamefont {Horiuchi},\ and\ \citenamefont {Tohsaki}}]{Zhou:2020nonlocalized}%
  \BibitemOpen
  \bibfield  {author} {\bibinfo {author} {\bibfnamefont {B.}~\bibnamefont {Zhou}}, \bibinfo {author} {\bibfnamefont {Y.}~\bibnamefont {Funaki}}, \bibinfo {author} {\bibfnamefont {H.}~\bibnamefont {Horiuchi}},\ and\ \bibinfo {author} {\bibfnamefont {A.}~\bibnamefont {Tohsaki}},\ }\href {https://doi.org/10.1007/s11467-019-0917-0} {\bibfield  {journal} {\bibinfo  {journal} {Front. Phys.}\ }\textbf {\bibinfo {volume} {15}},\ \bibinfo {pages} {14401} (\bibinfo {year} {2020})}\BibitemShut {NoStop}%
\bibitem [{\citenamefont {Tohsaki}(1994)}]{Tohsaki:1994zz}%
  \BibitemOpen
  \bibfield  {author} {\bibinfo {author} {\bibfnamefont {A.}~\bibnamefont {Tohsaki}},\ }\href {https://doi.org/10.1103/PhysRevC.49.1814} {\bibfield  {journal} {\bibinfo  {journal} {Phys. Rev. C}\ }\textbf {\bibinfo {volume} {49}},\ \bibinfo {pages} {1814} (\bibinfo {year} {1994})}\BibitemShut {NoStop}%
\bibitem [{\citenamefont {{Kanada-En'yo}}\ and\ \citenamefont {Akaishi}(2004)}]{enyo:2004}%
  \BibitemOpen
  \bibfield  {author} {\bibinfo {author} {\bibfnamefont {Y.}~\bibnamefont {{Kanada-En'yo}}}\ and\ \bibinfo {author} {\bibfnamefont {Y.}~\bibnamefont {Akaishi}},\ }\href {https://doi.org/10.1103/PhysRevC.69.034306} {\bibfield  {journal} {\bibinfo  {journal} {Phys. Rev. C}\ }\textbf {\bibinfo {volume} {69}},\ \bibinfo {pages} {034306} (\bibinfo {year} {2004})}\BibitemShut {NoStop}%
\bibitem [{\citenamefont {Dufour}\ and\ \citenamefont {Descouvemont}(2005)}]{Dufour:2005boe}%
  \BibitemOpen
  \bibfield  {author} {\bibinfo {author} {\bibfnamefont {M.}~\bibnamefont {Dufour}}\ and\ \bibinfo {author} {\bibfnamefont {P.}~\bibnamefont {Descouvemont}},\ }\href {https://doi.org/10.1016/j.nuclphysa.2004.12.041} {\bibfield  {journal} {\bibinfo  {journal} {Nucl. Phys. A}\ }\textbf {\bibinfo {volume} {750}},\ \bibinfo {pages} {218} (\bibinfo {year} {2005})}\BibitemShut {NoStop}%
\bibitem [{\citenamefont {Itagaki}\ \emph {et~al.}()\citenamefont {Itagaki}, \citenamefont {Fukui},\ and\ \citenamefont {Tohsaki}}]{Itagaki:2020pib}%
  \BibitemOpen
  \bibfield  {author} {\bibinfo {author} {\bibfnamefont {N.}~\bibnamefont {Itagaki}}, \bibinfo {author} {\bibfnamefont {T.}~\bibnamefont {Fukui}},\ and\ \bibinfo {author} {\bibfnamefont {A.}~\bibnamefont {Tohsaki}},\ }\href {https://arxiv.org/abs/2003.08546} {\bibinfo  {journal} {arXiv:2003.08546}\ }\BibitemShut {NoStop}%
\bibitem [{\citenamefont {Irvine}\ \emph {et~al.}(1971)\citenamefont {Irvine}, \citenamefont {Latorre},\ and\ \citenamefont {Pucknell}}]{Irvine01111971}%
  \BibitemOpen
\bibfield  {journal} {  }\bibfield  {author} {\bibinfo {author} {\bibfnamefont {J.}~\bibnamefont {Irvine}}, \bibinfo {author} {\bibfnamefont {C.}~\bibnamefont {Latorre}},\ and\ \bibinfo {author} {\bibfnamefont {V.}~\bibnamefont {Pucknell}},\ }\href {https://doi.org/10.1080/00018737100101321} {\bibfield  {journal} {\bibinfo  {journal} {Adv. Phys.}\ }\textbf {\bibinfo {volume} {20}},\ \bibinfo {pages} {661} (\bibinfo {year} {1971})}\BibitemShut {NoStop}%
\bibitem [{\citenamefont {Mitchell}\ \emph {et~al.}(1964)\citenamefont {Mitchell}, \citenamefont {Carter},\ and\ \citenamefont {Davis}}]{mitchell:1964}%
  \BibitemOpen
  \bibfield  {author} {\bibinfo {author} {\bibfnamefont {G.~E.}\ \bibnamefont {Mitchell}}, \bibinfo {author} {\bibfnamefont {E.~B.}\ \bibnamefont {Carter}},\ and\ \bibinfo {author} {\bibfnamefont {R.~H.}\ \bibnamefont {Davis}},\ }\href {https://doi.org/10.1103/PhysRev.133.B1434} {\bibfield  {journal} {\bibinfo  {journal} {Phys. Rev.}\ }\textbf {\bibinfo {volume} {133}},\ \bibinfo {pages} {B1434} (\bibinfo {year} {1964})}\BibitemShut {NoStop}%
\bibitem [{\citenamefont {Itagaki}\ \emph {et~al.}(1995)\citenamefont {Itagaki}, \citenamefont {Ohnishi},\ and\ \citenamefont {Katō}}]{Itagaki:1995}%
  \BibitemOpen
  \bibfield  {author} {\bibinfo {author} {\bibfnamefont {N.}~\bibnamefont {Itagaki}}, \bibinfo {author} {\bibfnamefont {A.}~\bibnamefont {Ohnishi}},\ and\ \bibinfo {author} {\bibfnamefont {K.}~\bibnamefont {Katō}},\ }\href {https://doi.org/10.1143/PTP.94.1019} {\bibfield  {journal} {\bibinfo  {journal} {Prog. Theor. Phys.}\ }\textbf {\bibinfo {volume} {94}},\ \bibinfo {pages} {1019} (\bibinfo {year} {1995})}\BibitemShut {NoStop}%
\bibitem [{\citenamefont {Itagaki}(2016)}]{Itagaki:2016bxb}%
  \BibitemOpen
  \bibfield  {author} {\bibinfo {author} {\bibfnamefont {N.}~\bibnamefont {Itagaki}},\ }\href {https://doi.org/10.1103/PhysRevC.94.064324} {\bibfield  {journal} {\bibinfo  {journal} {Phys. Rev. C}\ }\textbf {\bibinfo {volume} {94}},\ \bibinfo {pages} {064324} (\bibinfo {year} {2016})}\BibitemShut {NoStop}%
\bibitem [{\citenamefont {Hammer}\ \emph {et~al.}(2020)\citenamefont {Hammer}, \citenamefont {K{\"o}nig},\ and\ \citenamefont {van Kolck}}]{Hammer:2019poc}%
  \BibitemOpen
  \bibfield  {author} {\bibinfo {author} {\bibfnamefont {H.-W.}\ \bibnamefont {Hammer}}, \bibinfo {author} {\bibfnamefont {S.}~\bibnamefont {K{\"o}nig}},\ and\ \bibinfo {author} {\bibfnamefont {U.}~\bibnamefont {van Kolck}},\ }\href {https://doi.org/10.1103/RevModPhys.92.025004} {\bibfield  {journal} {\bibinfo  {journal} {Rev. Mod. Phys.}\ }\textbf {\bibinfo {volume} {92}},\ \bibinfo {pages} {025004} (\bibinfo {year} {2020})}\BibitemShut {NoStop}%
\bibitem [{\citenamefont {Epelbaum}\ \emph {et~al.}(2020)\citenamefont {Epelbaum}, \citenamefont {Krebs},\ and\ \citenamefont {Reinert}}]{Epelbaum:2020fip}%
  \BibitemOpen
  \bibfield  {author} {\bibinfo {author} {\bibfnamefont {E.}~\bibnamefont {Epelbaum}}, \bibinfo {author} {\bibfnamefont {H.}~\bibnamefont {Krebs}},\ and\ \bibinfo {author} {\bibfnamefont {P.}~\bibnamefont {Reinert}},\ }\href {https://doi.org/10.3389/fphy.2020.00098} {\bibfield  {journal} {\bibinfo  {journal} {Front. Phys.}\ }\textbf {\bibinfo {volume} {8}},\ \bibinfo {pages} {98} (\bibinfo {year} {2020})}\BibitemShut {NoStop}%
\bibitem [{\citenamefont {Machleidt}\ and\ \citenamefont {Sammarruca}(2024)}]{MACHLEIDT:2024}%
  \BibitemOpen
  \bibfield  {author} {\bibinfo {author} {\bibfnamefont {R.}~\bibnamefont {Machleidt}}\ and\ \bibinfo {author} {\bibfnamefont {F.}~\bibnamefont {Sammarruca}},\ }\href {https://doi.org/https://doi.org/10.1016/j.ppnp.2024.104117} {\bibfield  {journal} {\bibinfo  {journal} {Prog. Part. Nucl. Phys.}\ }\textbf {\bibinfo {volume} {137}},\ \bibinfo {pages} {104117} (\bibinfo {year} {2024})}\BibitemShut {NoStop}%
\bibitem [{\citenamefont {Fukui}(2022)}]{Fukui:2020ylj}%
  \BibitemOpen
  \bibfield  {author} {\bibinfo {author} {\bibfnamefont {T.}~\bibnamefont {Fukui}},\ }\href {https://doi.org/10.1088/1361-6471/ac58b3} {\bibfield  {journal} {\bibinfo  {journal} {J. Phys. G}\ }\textbf {\bibinfo {volume} {49}},\ \bibinfo {pages} {055102} (\bibinfo {year} {2022})}\BibitemShut {NoStop}%
\bibitem [{\citenamefont {Wu}\ \emph {et~al.}(2026)\citenamefont {Wu}, \citenamefont {Elhatisari}, \citenamefont {Mei\ss{}ner}, \citenamefont {Shen}, \citenamefont {Geng},\ and\ \citenamefont {Kim}}]{wu2026search}%
  \BibitemOpen
  \bibfield  {author} {\bibinfo {author} {\bibfnamefont {L.}~\bibnamefont {Wu}}, \bibinfo {author} {\bibfnamefont {S.}~\bibnamefont {Elhatisari}}, \bibinfo {author} {\bibfnamefont {U.-G.}\ \bibnamefont {Mei\ss{}ner}}, \bibinfo {author} {\bibfnamefont {S.}~\bibnamefont {Shen}}, \bibinfo {author} {\bibfnamefont {L.-S.}\ \bibnamefont {Geng}},\ and\ \bibinfo {author} {\bibfnamefont {Y.}~\bibnamefont {Kim}},\ }\href {https://doi.org/10.1103/89w9-p443} {\bibfield  {journal} {\bibinfo  {journal} {Phys. Rev. Lett.}\ }\textbf {\bibinfo {volume} {136}},\ \bibinfo {pages} {142502} (\bibinfo {year} {2026})}\BibitemShut {NoStop}%
\bibitem [{\citenamefont {Wigner}(1937)}]{Wigner:1936dx}%
  \BibitemOpen
  \bibfield  {author} {\bibinfo {author} {\bibfnamefont {E.}~\bibnamefont {Wigner}},\ }\href {https://doi.org/10.1103/PhysRev.51.106} {\bibfield  {journal} {\bibinfo  {journal} {Phys. Rev.}\ }\textbf {\bibinfo {volume} {51}},\ \bibinfo {pages} {106} (\bibinfo {year} {1937})}\BibitemShut {NoStop}%
\bibitem [{\citenamefont {Kaplan}\ and\ \citenamefont {Savage}(1996)}]{Kaplan:1995yg}%
  \BibitemOpen
  \bibfield  {author} {\bibinfo {author} {\bibfnamefont {D.~B.}\ \bibnamefont {Kaplan}}\ and\ \bibinfo {author} {\bibfnamefont {M.~J.}\ \bibnamefont {Savage}},\ }\href {https://doi.org/10.1016/0370-2693(95)01277-X} {\bibfield  {journal} {\bibinfo  {journal} {Phys. Lett. B}\ }\textbf {\bibinfo {volume} {365}},\ \bibinfo {pages} {244} (\bibinfo {year} {1996})}\BibitemShut {NoStop}%
\bibitem [{\citenamefont {Kaplan}\ and\ \citenamefont {Manohar}(1997)}]{Kaplan:1996rk}%
  \BibitemOpen
  \bibfield  {author} {\bibinfo {author} {\bibfnamefont {D.~B.}\ \bibnamefont {Kaplan}}\ and\ \bibinfo {author} {\bibfnamefont {A.~V.}\ \bibnamefont {Manohar}},\ }\href {https://doi.org/10.1103/PhysRevC.56.76} {\bibfield  {journal} {\bibinfo  {journal} {Phys. Rev. C}\ }\textbf {\bibinfo {volume} {56}},\ \bibinfo {pages} {76} (\bibinfo {year} {1997})}\BibitemShut {NoStop}%
\bibitem [{\citenamefont {Lu}\ \emph {et~al.}(2019)\citenamefont {Lu}, \citenamefont {Li}, \citenamefont {Elhatisari}, \citenamefont {Lee}, \citenamefont {Epelbaum},\ and\ \citenamefont {Mei{\ss}ner}}]{Lu:2018bat}%
  \BibitemOpen
  \bibfield  {author} {\bibinfo {author} {\bibfnamefont {B.-N.}\ \bibnamefont {Lu}}, \bibinfo {author} {\bibfnamefont {N.}~\bibnamefont {Li}}, \bibinfo {author} {\bibfnamefont {S.}~\bibnamefont {Elhatisari}}, \bibinfo {author} {\bibfnamefont {D.}~\bibnamefont {Lee}}, \bibinfo {author} {\bibfnamefont {E.}~\bibnamefont {Epelbaum}},\ and\ \bibinfo {author} {\bibfnamefont {U.-G.}\ \bibnamefont {Mei{\ss}ner}},\ }\href {https://doi.org/10.1016/j.physletb.2019.134863} {\bibfield  {journal} {\bibinfo  {journal} {Phys. Lett. B}\ }\textbf {\bibinfo {volume} {797}},\ \bibinfo {pages} {134863} (\bibinfo {year} {2019})}\BibitemShut {NoStop}%
\bibitem [{\citenamefont {Niu}\ and\ \citenamefont {Lu}(2025)}]{niu:2025spf}%
  \BibitemOpen
  \bibfield  {author} {\bibinfo {author} {\bibfnamefont {Z.-W.}\ \bibnamefont {Niu}}\ and\ \bibinfo {author} {\bibfnamefont {B.-N.}\ \bibnamefont {Lu}},\ }\href {https://doi.org/10.1103/pn99-6dxt} {\bibfield  {journal} {\bibinfo  {journal} {Phys. Rev. Lett.}\ }\textbf {\bibinfo {volume} {135}},\ \bibinfo {pages} {222504} (\bibinfo {year} {2025})}\BibitemShut {NoStop}%
\bibitem [{\citenamefont {Shen}\ \emph {et~al.}(2023)\citenamefont {Shen}, \citenamefont {Elhatisari}, \citenamefont {L{\"a}hde}, \citenamefont {Lee}, \citenamefont {Lu},\ and\ \citenamefont {Mei{\ss}ner}}]{Shen:2022bak}%
  \BibitemOpen
  \bibfield  {author} {\bibinfo {author} {\bibfnamefont {S.}~\bibnamefont {Shen}}, \bibinfo {author} {\bibfnamefont {S.}~\bibnamefont {Elhatisari}}, \bibinfo {author} {\bibfnamefont {T.~A.}\ \bibnamefont {L{\"a}hde}}, \bibinfo {author} {\bibfnamefont {D.}~\bibnamefont {Lee}}, \bibinfo {author} {\bibfnamefont {B.-N.}\ \bibnamefont {Lu}},\ and\ \bibinfo {author} {\bibfnamefont {U.-G.}\ \bibnamefont {Mei{\ss}ner}},\ }\href {https://doi.org/10.1038/s41467-023-38391-y} {\bibfield  {journal} {\bibinfo  {journal} {Nat. Commun.}\ }\textbf {\bibinfo {volume} {14}},\ \bibinfo {pages} {2777} (\bibinfo {year} {2023})}\BibitemShut {NoStop}%
\bibitem [{\citenamefont {Shen}\ \emph {et~al.}(2025)\citenamefont {Shen}, \citenamefont {Elhatisari}, \citenamefont {Lee}, \citenamefont {Mei{\ss}ner},\ and\ \citenamefont {Ren}}]{Shen:2024qzi}%
  \BibitemOpen
  \bibfield  {author} {\bibinfo {author} {\bibfnamefont {S.}~\bibnamefont {Shen}}, \bibinfo {author} {\bibfnamefont {S.}~\bibnamefont {Elhatisari}}, \bibinfo {author} {\bibfnamefont {D.}~\bibnamefont {Lee}}, \bibinfo {author} {\bibfnamefont {U.-G.}\ \bibnamefont {Mei{\ss}ner}},\ and\ \bibinfo {author} {\bibfnamefont {Z.}~\bibnamefont {Ren}},\ }\href {https://doi.org/10.1103/PhysRevLett.134.162503} {\bibfield  {journal} {\bibinfo  {journal} {Phys. Rev. Lett.}\ }\textbf {\bibinfo {volume} {134}},\ \bibinfo {pages} {162503} (\bibinfo {year} {2025})}\BibitemShut {NoStop}%
\bibitem [{\citenamefont {Lyu}\ \emph {et~al.}()\citenamefont {Lyu}, \citenamefont {Zuo}, \citenamefont {Peng}, \citenamefont {K\"onig},\ and\ \citenamefont {Long}}]{Lyu:2025yhz}%
  \BibitemOpen
  \bibfield  {author} {\bibinfo {author} {\bibfnamefont {S.}~\bibnamefont {Lyu}}, \bibinfo {author} {\bibfnamefont {L.}~\bibnamefont {Zuo}}, \bibinfo {author} {\bibfnamefont {R.}~\bibnamefont {Peng}}, \bibinfo {author} {\bibfnamefont {S.}~\bibnamefont {K\"onig}},\ and\ \bibinfo {author} {\bibfnamefont {B.}~\bibnamefont {Long}},\ }\href {https://arxiv.org/abs/2511.12522} {\bibinfo  {journal} {arXiv:2511.12522}\ }\BibitemShut {NoStop}%
\bibitem [{\citenamefont {Brink}(1966)}]{brink1966alpha}%
  \BibitemOpen
\bibfield  {journal} {  }\bibfield  {author} {\bibinfo {author} {\bibfnamefont {D.~M.}\ \bibnamefont {Brink}},\ }\href@noop {} {\emph {\bibinfo {title} {The Alpha-Particle Model of Light Nuclei}}}\ (\bibinfo  {publisher} {Academic Press},\ \bibinfo {address} {New York},\ \bibinfo {year} {1966})\BibitemShut {NoStop}%
\bibitem [{\citenamefont {Hill}\ and\ \citenamefont {Wheeler}(1953)}]{Hill:1952jb}%
  \BibitemOpen
  \bibfield  {author} {\bibinfo {author} {\bibfnamefont {D.~L.}\ \bibnamefont {Hill}}\ and\ \bibinfo {author} {\bibfnamefont {J.~A.}\ \bibnamefont {Wheeler}},\ }\href {https://doi.org/10.1103/PhysRev.89.1102} {\bibfield  {journal} {\bibinfo  {journal} {Phys. Rev.}\ }\textbf {\bibinfo {volume} {89}},\ \bibinfo {pages} {1102} (\bibinfo {year} {1953})}\BibitemShut {NoStop}%
\bibitem [{\citenamefont {Phillips}\ and\ \citenamefont {Schat}(2013)}]{Phillips:2013rsa}%
  \BibitemOpen
  \bibfield  {author} {\bibinfo {author} {\bibfnamefont {D.~R.}\ \bibnamefont {Phillips}}\ and\ \bibinfo {author} {\bibfnamefont {C.}~\bibnamefont {Schat}},\ }\href {https://doi.org/10.1103/PhysRevC.88.034002} {\bibfield  {journal} {\bibinfo  {journal} {Phys. Rev. C}\ }\textbf {\bibinfo {volume} {88}},\ \bibinfo {pages} {034002} (\bibinfo {year} {2013})}\BibitemShut {NoStop}%
\bibitem [{\citenamefont {Epelbaum}\ \emph {et~al.}(2015)\citenamefont {Epelbaum}, \citenamefont {Gasparyan}, \citenamefont {Krebs},\ and\ \citenamefont {Schat}}]{Epelbaum:2014sea}%
  \BibitemOpen
  \bibfield  {author} {\bibinfo {author} {\bibfnamefont {E.}~\bibnamefont {Epelbaum}}, \bibinfo {author} {\bibfnamefont {A.~M.}\ \bibnamefont {Gasparyan}}, \bibinfo {author} {\bibfnamefont {H.}~\bibnamefont {Krebs}},\ and\ \bibinfo {author} {\bibfnamefont {C.}~\bibnamefont {Schat}},\ }\href {https://doi.org/10.1140/epja/i2015-15026-y} {\bibfield  {journal} {\bibinfo  {journal} {Eur. Phys. J. A}\ }\textbf {\bibinfo {volume} {51}},\ \bibinfo {pages} {26} (\bibinfo {year} {2015})}\BibitemShut {NoStop}%
\bibitem [{\citenamefont {Li}\ \emph {et~al.}(2026{\natexlab{b}})\citenamefont {Li}, \citenamefont {Zhou}, \citenamefont {Bai}, \citenamefont {Zhou},\ and\ \citenamefont {Ma}}]{supplement}%
  \BibitemOpen
  \bibfield  {author} {\bibinfo {author} {\bibfnamefont {G.-P.}\ \bibnamefont {Li}}, \bibinfo {author} {\bibfnamefont {S.-Y.}\ \bibnamefont {Zhou}}, \bibinfo {author} {\bibfnamefont {D.}~\bibnamefont {Bai}}, \bibinfo {author} {\bibfnamefont {B.}~\bibnamefont {Zhou}},\ and\ \bibinfo {author} {\bibfnamefont {Y.-G.}\ \bibnamefont {Ma}},\ }\href@noop {} {\bibinfo {title} {Supplemental material for ``minimal wigner-$\ensuremath{SU(4)}$ interaction in microscopic cluster models for $\alpha$-conjugate nuclei''}} (\bibinfo {year} {2026}{\natexlab{b}}),\ \bibinfo {note} {see Supplemental Material for details on the parameter determination, cutoff dependence, basis construction, and additional results for $^{20}\mathrm{Ne}$.}\BibitemShut {Stop}%
\bibitem [{\citenamefont {Mito}\ and\ \citenamefont {Kamimura}(1976)}]{Mito1976}%
  \BibitemOpen
  \bibfield  {author} {\bibinfo {author} {\bibfnamefont {Y.}~\bibnamefont {Mito}}\ and\ \bibinfo {author} {\bibfnamefont {M.}~\bibnamefont {Kamimura}},\ }\href {https://doi.org/10.1143/PTP.56.583} {\bibfield  {journal} {\bibinfo  {journal} {Prog. Theor. Phys.}\ }\textbf {\bibinfo {volume} {56}},\ \bibinfo {pages} {583} (\bibinfo {year} {1976})}\BibitemShut {NoStop}%
\bibitem [{\citenamefont {Kamimura}(1977)}]{Kamimura:1977okl}%
  \BibitemOpen
  \bibfield  {author} {\bibinfo {author} {\bibfnamefont {M.}~\bibnamefont {Kamimura}},\ }\href {https://doi.org/10.1143/PTPS.62.236} {\bibfield  {journal} {\bibinfo  {journal} {Prog. Theor. Phys. Suppl.}\ }\textbf {\bibinfo {volume} {62}},\ \bibinfo {pages} {236} (\bibinfo {year} {1977})}\BibitemShut {NoStop}%
\bibitem [{\citenamefont {Lin}\ \emph {et~al.}(2022)\citenamefont {Lin}, \citenamefont {Hammer},\ and\ \citenamefont {Mei{\ss}ner}}]{Lin:2021xrc}%
  \BibitemOpen
  \bibfield  {author} {\bibinfo {author} {\bibfnamefont {Y.-H.}\ \bibnamefont {Lin}}, \bibinfo {author} {\bibfnamefont {H.-W.}\ \bibnamefont {Hammer}},\ and\ \bibinfo {author} {\bibfnamefont {U.-G.}\ \bibnamefont {Mei{\ss}ner}},\ }\href {https://doi.org/10.1103/PhysRevLett.128.052002} {\bibfield  {journal} {\bibinfo  {journal} {Phys. Rev. Lett.}\ }\textbf {\bibinfo {volume} {128}},\ \bibinfo {pages} {052002} (\bibinfo {year} {2022})}\BibitemShut {NoStop}%
\bibitem [{\citenamefont {{Ajzenberg-Selove}}(1990)}]{Ajzenberg-Selove:1990fsm}%
  \BibitemOpen
  \bibfield  {author} {\bibinfo {author} {\bibfnamefont {F.}~\bibnamefont {{Ajzenberg-Selove}}},\ }\href {https://doi.org/10.1016/0375-9474(90)90271-M} {\bibfield  {journal} {\bibinfo  {journal} {Nucl. Phys. A}\ }\textbf {\bibinfo {volume} {506}},\ \bibinfo {pages} {1} (\bibinfo {year} {1990})}\BibitemShut {NoStop}%
\bibitem [{\citenamefont {Angeli}\ and\ \citenamefont {Marinova}(2013)}]{Angeli:2013epw}%
  \BibitemOpen
  \bibfield  {author} {\bibinfo {author} {\bibfnamefont {I.}~\bibnamefont {Angeli}}\ and\ \bibinfo {author} {\bibfnamefont {K.~P.}\ \bibnamefont {Marinova}},\ }\href {https://doi.org/10.1016/j.adt.2011.12.006} {\bibfield  {journal} {\bibinfo  {journal} {At. Data Nucl. Data Tabl.}\ }\textbf {\bibinfo {volume} {99}},\ \bibinfo {pages} {69} (\bibinfo {year} {2013})}\BibitemShut {NoStop}%
\bibitem [{\citenamefont {Kelley}\ \emph {et~al.}(2017)\citenamefont {Kelley}, \citenamefont {Purcell},\ and\ \citenamefont {Sheu}}]{Kelley:2017qgh}%
  \BibitemOpen
  \bibfield  {author} {\bibinfo {author} {\bibfnamefont {J.}~\bibnamefont {Kelley}}, \bibinfo {author} {\bibfnamefont {J.}~\bibnamefont {Purcell}},\ and\ \bibinfo {author} {\bibfnamefont {C.}~\bibnamefont {Sheu}},\ }\href {https://doi.org/10.1016/j.nuclphysa.2017.07.015} {\bibfield  {journal} {\bibinfo  {journal} {Nucl. Phys. A}\ }\textbf {\bibinfo {volume} {968}},\ \bibinfo {pages} {71} (\bibinfo {year} {2017})}\BibitemShut {NoStop}%
\bibitem [{\citenamefont {Imai}\ \emph {et~al.}(2019)\citenamefont {Imai}, \citenamefont {Tada},\ and\ \citenamefont {Kimura}}]{Imai:2018lww}%
  \BibitemOpen
  \bibfield  {author} {\bibinfo {author} {\bibfnamefont {R.}~\bibnamefont {Imai}}, \bibinfo {author} {\bibfnamefont {T.}~\bibnamefont {Tada}},\ and\ \bibinfo {author} {\bibfnamefont {M.}~\bibnamefont {Kimura}},\ }\href {https://doi.org/10.1103/PhysRevC.99.064327} {\bibfield  {journal} {\bibinfo  {journal} {Phys. Rev. C}\ }\textbf {\bibinfo {volume} {99}},\ \bibinfo {pages} {064327} (\bibinfo {year} {2019})}\BibitemShut {NoStop}%
\bibitem [{\citenamefont {Funaki}(2015)}]{Funaki:2014tda}%
  \BibitemOpen
  \bibfield  {author} {\bibinfo {author} {\bibfnamefont {Y.}~\bibnamefont {Funaki}},\ }\href {https://doi.org/10.1103/PhysRevC.92.021302} {\bibfield  {journal} {\bibinfo  {journal} {Phys. Rev. C}\ }\textbf {\bibinfo {volume} {92}},\ \bibinfo {pages} {021302} (\bibinfo {year} {2015})}\BibitemShut {NoStop}%
\bibitem [{\citenamefont {Funaki}(2016)}]{Funaki:2016atc}%
  \BibitemOpen
  \bibfield  {author} {\bibinfo {author} {\bibfnamefont {Y.}~\bibnamefont {Funaki}},\ }\href {https://doi.org/10.1103/PhysRevC.94.024344} {\bibfield  {journal} {\bibinfo  {journal} {Phys. Rev. C}\ }\textbf {\bibinfo {volume} {94}},\ \bibinfo {pages} {024344} (\bibinfo {year} {2016})}\BibitemShut {NoStop}%
\bibitem [{\citenamefont {deBoer}\ \emph {et~al.}(2017)\citenamefont {deBoer}, \citenamefont {G\"orres}, \citenamefont {Wiescher}, \citenamefont {Azuma}, \citenamefont {Best}, \citenamefont {Brune}, \citenamefont {Fields}, \citenamefont {Jones}, \citenamefont {Pignatari}, \citenamefont {Sayre}, \citenamefont {Smith}, \citenamefont {Timmes},\ and\ \citenamefont {Uberseder}}]{deboer:2017rmp}%
  \BibitemOpen
  \bibfield  {author} {\bibinfo {author} {\bibfnamefont {R.~J.}\ \bibnamefont {deBoer}}, \bibinfo {author} {\bibfnamefont {J.}~\bibnamefont {G\"orres}}, \bibinfo {author} {\bibfnamefont {M.}~\bibnamefont {Wiescher}}, \bibinfo {author} {\bibfnamefont {R.~E.}\ \bibnamefont {Azuma}}, \bibinfo {author} {\bibfnamefont {A.}~\bibnamefont {Best}}, \bibinfo {author} {\bibfnamefont {C.~R.}\ \bibnamefont {Brune}}, \bibinfo {author} {\bibfnamefont {C.~E.}\ \bibnamefont {Fields}}, \bibinfo {author} {\bibfnamefont {S.}~\bibnamefont {Jones}}, \bibinfo {author} {\bibfnamefont {M.}~\bibnamefont {Pignatari}}, \bibinfo {author} {\bibfnamefont {D.}~\bibnamefont {Sayre}}, \bibinfo {author} {\bibfnamefont {K.}~\bibnamefont {Smith}}, \bibinfo {author} {\bibfnamefont {F.~X.}\ \bibnamefont {Timmes}},\ and\ \bibinfo {author} {\bibfnamefont {E.}~\bibnamefont {Uberseder}},\ }\href {https://doi.org/10.1103/RevModPhys.89.035007} {\bibfield  {journal} {\bibinfo  {journal} {Rev. Mod. Phys.}\ }\textbf {\bibinfo {volume} {89}},\ \bibinfo {pages} {035007} (\bibinfo {year} {2017})}\BibitemShut {NoStop}%
\bibitem [{\citenamefont {Jiang}\ \emph {et~al.}(2007)\citenamefont {Jiang}, \citenamefont {Rehm}, \citenamefont {Back},\ and\ \citenamefont {Janssens}}]{cljiang:2007fusion}%
  \BibitemOpen
  \bibfield  {author} {\bibinfo {author} {\bibfnamefont {C.~L.}\ \bibnamefont {Jiang}}, \bibinfo {author} {\bibfnamefont {K.~E.}\ \bibnamefont {Rehm}}, \bibinfo {author} {\bibfnamefont {B.~B.}\ \bibnamefont {Back}},\ and\ \bibinfo {author} {\bibfnamefont {R.~V.~F.}\ \bibnamefont {Janssens}},\ }\href {https://doi.org/10.1103/PhysRevC.75.015803} {\bibfield  {journal} {\bibinfo  {journal} {Phys. Rev. C}\ }\textbf {\bibinfo {volume} {75}},\ \bibinfo {pages} {015803} (\bibinfo {year} {2007})}\BibitemShut {NoStop}%
\bibitem [{\citenamefont {Li}\ \emph {et~al.}(2020)\citenamefont {Li}, \citenamefont {Zhang},\ and\ \citenamefont {Ma}}]{Li2020}%
  \BibitemOpen
  \bibfield  {author} {\bibinfo {author} {\bibfnamefont {Y.-A.}\ \bibnamefont {Li}}, \bibinfo {author} {\bibfnamefont {S.}~\bibnamefont {Zhang}},\ and\ \bibinfo {author} {\bibfnamefont {Y.-G.}\ \bibnamefont {Ma}},\ }\href {https://doi.org/10.1103/PhysRevC.102.054907} {\bibfield  {journal} {\bibinfo  {journal} {Phys. Rev. C}\ }\textbf {\bibinfo {volume} {102}},\ \bibinfo {pages} {054907} (\bibinfo {year} {2020})}\BibitemShut {NoStop}%
\bibitem [{\citenamefont {Abualrob}\ \emph {et~al.}(2026)\citenamefont {Abualrob} \emph {et~al.}}]{ALICE:2026geometryOO}%
  \BibitemOpen
  \bibfield  {author} {\bibinfo {author} {\bibfnamefont {I.~J.}\ \bibnamefont {Abualrob}} \emph {et~al.} (\bibinfo {collaboration} {ALICE Collaboration}),\ }\href {https://link.aps.org/doi/10.1103/gymp-vp87} {\bibfield  {journal} {\bibinfo  {journal} {Phys. Rev. Lett.}\ } (\bibinfo {year} {2026})}\BibitemShut {NoStop}%
\bibitem [{\citenamefont {Hayrapetyan}\ \emph {et~al.}(2026)\citenamefont {Hayrapetyan} \emph {et~al.}}]{CMS:2026OONeNeFlow}%
  \BibitemOpen
  \bibfield  {author} {\bibinfo {author} {\bibfnamefont {A.}~\bibnamefont {Hayrapetyan}} \emph {et~al.} (\bibinfo {collaboration} {CMS Collaboration}),\ }\href {https://link.aps.org/doi/10.1103/26wx-tg6f} {\bibfield  {journal} {\bibinfo  {journal} {Phys. Rev. Lett.}\ } (\bibinfo {year} {2026})}\BibitemShut {NoStop}%
\bibitem [{\citenamefont {Chen}\ \emph {et~al.}(2023)\citenamefont {Chen}, \citenamefont {Ye}, \citenamefont {Ma}, \citenamefont {Han}, \citenamefont {Wang}, \citenamefont {Lin}, \citenamefont {Jia}, \citenamefont {Yang}, \citenamefont {Li}, \citenamefont {Yang}, \citenamefont {Hu}, \citenamefont {Tan}, \citenamefont {Wei}, \citenamefont {Pu}, \citenamefont {Chen}, \citenamefont {Lou}, \citenamefont {Yang}, \citenamefont {Li}, \citenamefont {Yang}, \citenamefont {Luo}, \citenamefont {Huang}, \citenamefont {Zhong}, \citenamefont {Li},\ and\ \citenamefont {Xu}}]{chen_new_2023-1}%
  \BibitemOpen
  \bibfield  {author} {\bibinfo {author} {\bibfnamefont {J.}~\bibnamefont {Chen}}, \bibinfo {author} {\bibfnamefont {Y.}~\bibnamefont {Ye}}, \bibinfo {author} {\bibfnamefont {K.}~\bibnamefont {Ma}}, \bibinfo {author} {\bibfnamefont {J.}~\bibnamefont {Han}}, \bibinfo {author} {\bibfnamefont {D.}~\bibnamefont {Wang}}, \bibinfo {author} {\bibfnamefont {C.}~\bibnamefont {Lin}}, \bibinfo {author} {\bibfnamefont {H.}~\bibnamefont {Jia}}, \bibinfo {author} {\bibfnamefont {L.}~\bibnamefont {Yang}}, \bibinfo {author} {\bibfnamefont {G.}~\bibnamefont {Li}}, \bibinfo {author} {\bibfnamefont {L.}~\bibnamefont {Yang}}, \bibinfo {author} {\bibfnamefont {Z.}~\bibnamefont {Hu}}, \bibinfo {author} {\bibfnamefont {Z.}~\bibnamefont {Tan}}, \bibinfo {author} {\bibfnamefont {K.}~\bibnamefont {Wei}}, \bibinfo {author} {\bibfnamefont {W.-L.}\ \bibnamefont {Pu}}, \bibinfo {author} {\bibfnamefont {Y.}~\bibnamefont {Chen}}, \bibinfo {author} {\bibfnamefont {J.}~\bibnamefont {Lou}}, \bibinfo {author} {\bibfnamefont {X.}~\bibnamefont {Yang}}, \bibinfo {author} {\bibfnamefont {Q.}~\bibnamefont {Li}}, \bibinfo {author} {\bibfnamefont {Z.}~\bibnamefont {Yang}}, \bibinfo {author} {\bibfnamefont {T.}~\bibnamefont {Luo}}, \bibinfo {author} {\bibfnamefont {D.}~\bibnamefont {Huang}}, \bibinfo {author} {\bibfnamefont {S.}~\bibnamefont {Zhong}}, \bibinfo {author} {\bibfnamefont {Z.}~\bibnamefont {Li}},\ and\ \bibinfo {author} {\bibfnamefont {J.}~\bibnamefont {Xu}},\ }\href {https://doi.org/10.1016/j.scib.2023.04.031} {\bibfield  {journal} {\bibinfo  {journal} {Sci. Bull.}\ }\textbf {\bibinfo {volume} {68}},\ \bibinfo {pages} {1119} (\bibinfo {year} {2023})}\BibitemShut {NoStop}%
\bibitem [{\citenamefont {Lee}(2007)}]{lee:2007sca}%
  \BibitemOpen
  \bibfield  {author} {\bibinfo {author} {\bibfnamefont {D.}~\bibnamefont {Lee}},\ }\href {https://doi.org/10.1103/PhysRevLett.98.182501} {\bibfield  {journal} {\bibinfo  {journal} {Phys. Rev. Lett.}\ }\textbf {\bibinfo {volume} {98}},\ \bibinfo {pages} {182501} (\bibinfo {year} {2007})}\BibitemShut {NoStop}%
\bibitem [{\citenamefont {Matsuse}\ \emph {et~al.}(1975)\citenamefont {Matsuse}, \citenamefont {Kamimura},\ and\ \citenamefont {Fukushima}}]{matsuse_study_1975}%
  \BibitemOpen
  \bibfield  {author} {\bibinfo {author} {\bibfnamefont {T.}~\bibnamefont {Matsuse}}, \bibinfo {author} {\bibfnamefont {M.}~\bibnamefont {Kamimura}},\ and\ \bibinfo {author} {\bibfnamefont {Y.}~\bibnamefont {Fukushima}},\ }\href {https://doi.org/10.1143/PTP.53.706} {\bibfield  {journal} {\bibinfo  {journal} {Prog. Theor. Phys.}\ }\textbf {\bibinfo {volume} {53}},\ \bibinfo {pages} {706} (\bibinfo {year} {1975})}\BibitemShut {NoStop}%
\bibitem [{\citenamefont {Zhou}\ \emph {et~al.}(2013)\citenamefont {Zhou}, \citenamefont {Funaki}, \citenamefont {Horiuchi}, \citenamefont {Ren}, \citenamefont {R\"opke}, \citenamefont {Schuck}, \citenamefont {Tohsaki}, \citenamefont {Xu},\ and\ \citenamefont {Yamada}}]{zhou_nonlocalized_2013}%
  \BibitemOpen
  \bibfield  {author} {\bibinfo {author} {\bibfnamefont {B.}~\bibnamefont {Zhou}}, \bibinfo {author} {\bibfnamefont {Y.}~\bibnamefont {Funaki}}, \bibinfo {author} {\bibfnamefont {H.}~\bibnamefont {Horiuchi}}, \bibinfo {author} {\bibfnamefont {Z.}~\bibnamefont {Ren}}, \bibinfo {author} {\bibfnamefont {G.}~\bibnamefont {R\"opke}}, \bibinfo {author} {\bibfnamefont {P.}~\bibnamefont {Schuck}}, \bibinfo {author} {\bibfnamefont {A.}~\bibnamefont {Tohsaki}}, \bibinfo {author} {\bibfnamefont {C.}~\bibnamefont {Xu}},\ and\ \bibinfo {author} {\bibfnamefont {T.}~\bibnamefont {Yamada}},\ }\href {https://doi.org/10.1103/PhysRevLett.110.262501} {\bibfield  {journal} {\bibinfo  {journal} {Phys. Rev. Lett.}\ }\textbf {\bibinfo {volume} {110}},\ \bibinfo {pages} {262501} (\bibinfo {year} {2013})}\BibitemShut {NoStop}%
\bibitem [{\citenamefont {{Ajzenberg-Selove}}(1982)}]{Ajzenberg-Selove:1982fgy}%
  \BibitemOpen
  \bibfield  {author} {\bibinfo {author} {\bibfnamefont {F.}~\bibnamefont {{Ajzenberg-Selove}}},\ }\href {https://doi.org/10.1016/0375-9474(82)90538-3} {\bibfield  {journal} {\bibinfo  {journal} {Nucl. Phys. A}\ }\textbf {\bibinfo {volume} {375}},\ \bibinfo {pages} {1} (\bibinfo {year} {1982})}\BibitemShut {NoStop}%
\bibitem [{\citenamefont {Wakasa}\ \emph {et~al.}(2007)\citenamefont {Wakasa}, \citenamefont {Ihara}, \citenamefont {Fujita}, \citenamefont {Funaki}, \citenamefont {Hatanaka}, \citenamefont {Horiuchi}, \citenamefont {Itoh}, \citenamefont {Kamiya}, \citenamefont {R{\"o}pke}, \citenamefont {Sakaguchi}, \citenamefont {Sakamoto}, \citenamefont {Sakemi}, \citenamefont {Schuck}, \citenamefont {Shimizu}, \citenamefont {Takashina}, \citenamefont {Terashima}, \citenamefont {Tohsaki}, \citenamefont {Uchida}, \citenamefont {Yoshida},\ and\ \citenamefont {Yosoi}}]{Wakasa:2006nt}%
  \BibitemOpen
  \bibfield  {author} {\bibinfo {author} {\bibfnamefont {T.}~\bibnamefont {Wakasa}}, \bibinfo {author} {\bibfnamefont {E.}~\bibnamefont {Ihara}}, \bibinfo {author} {\bibfnamefont {K.}~\bibnamefont {Fujita}}, \bibinfo {author} {\bibfnamefont {Y.}~\bibnamefont {Funaki}}, \bibinfo {author} {\bibfnamefont {K.}~\bibnamefont {Hatanaka}}, \bibinfo {author} {\bibfnamefont {H.}~\bibnamefont {Horiuchi}}, \bibinfo {author} {\bibfnamefont {M.}~\bibnamefont {Itoh}}, \bibinfo {author} {\bibfnamefont {J.}~\bibnamefont {Kamiya}}, \bibinfo {author} {\bibfnamefont {G.}~\bibnamefont {R{\"o}pke}}, \bibinfo {author} {\bibfnamefont {H.}~\bibnamefont {Sakaguchi}}, \bibinfo {author} {\bibfnamefont {N.}~\bibnamefont {Sakamoto}}, \bibinfo {author} {\bibfnamefont {Y.}~\bibnamefont {Sakemi}}, \bibinfo {author} {\bibfnamefont {P.}~\bibnamefont {Schuck}}, \bibinfo {author} {\bibfnamefont {Y.}~\bibnamefont {Shimizu}}, \bibinfo {author} {\bibfnamefont {M.}~\bibnamefont {Takashina}}, \bibinfo {author} {\bibfnamefont {S.}~\bibnamefont {Terashima}}, \bibinfo {author} {\bibfnamefont {A.}~\bibnamefont {Tohsaki}}, \bibinfo {author} {\bibfnamefont {M.}~\bibnamefont {Uchida}}, \bibinfo {author} {\bibfnamefont {H.~P.}\ \bibnamefont {Yoshida}},\ and\ \bibinfo {author} {\bibfnamefont {M.}~\bibnamefont {Yosoi}},\ }\href {https://doi.org/10.1016/j.physletb.2007.08.016} {\bibfield  {journal} {\bibinfo  {journal} {Phys. Lett. B}\ }\textbf {\bibinfo {volume} {653}},\ \bibinfo {pages} {173} (\bibinfo {year} {2007})}\BibitemShut {NoStop}%
\bibitem [{\citenamefont {Zhou}\ \emph {et~al.}()\citenamefont {Zhou}, \citenamefont {Zhou}, \citenamefont {Itagaki},\ and\ \citenamefont {Tohsaki}}]{syzhou:2025}%
  \BibitemOpen
  \bibfield  {author} {\bibinfo {author} {\bibfnamefont {S.-Y.}\ \bibnamefont {Zhou}}, \bibinfo {author} {\bibfnamefont {B.}~\bibnamefont {Zhou}}, \bibinfo {author} {\bibfnamefont {N.}~\bibnamefont {Itagaki}},\ and\ \bibinfo {author} {\bibfnamefont {A.}~\bibnamefont {Tohsaki}},\ }\href@noop {} {\bibinfo  {journal} {under review (2026)}\ }\BibitemShut {NoStop}%
\bibitem [{\citenamefont {Dufour}\ and\ \citenamefont {Descouvemont}(2014)}]{Dufour:2014rta}%
  \BibitemOpen
\bibfield  {journal} {  }\bibfield  {author} {\bibinfo {author} {\bibfnamefont {M.}~\bibnamefont {Dufour}}\ and\ \bibinfo {author} {\bibfnamefont {P.}~\bibnamefont {Descouvemont}},\ }\href {https://doi.org/10.1016/j.nuclphysa.2014.04.010} {\bibfield  {journal} {\bibinfo  {journal} {Nucl. Phys. A}\ }\textbf {\bibinfo {volume} {927}},\ \bibinfo {pages} {134} (\bibinfo {year} {2014})}\BibitemShut {NoStop}%
\end{thebibliography}
\end{document}